# In-Vehicle PLC: In-Car and In-Ship Channel Characterization

Alberto Pittolo[(1)], Marco De Piante[(1)], Fabio Versolatto[(1)], and Andrea M. Tonello[(1-2)]
(1) University of Udine, Italy; (2) University of Klagenfurt, Austria

*This paper deals with power line communication (PLC) in the context of in-vehicle data networks. This technology can provide high-speed data connectivity via the exploitation of the existing power network, with clear potential benefits in terms of cost and weight reduction. The focus is on two scenarios: an electric car and a cruise ship. An overview of the wiring infrastructure and the network topology in these two scenarios is provided. The main findings reported in the literature related to the channel characteristics are reported. Noise is also assessed with emphasis to the electric car context. Then, new results from the statistical analysis of measurements made in a compact electric car and in a large cruise ship are shown. The channel characteristics are analysed in terms of average channel gain, delay spread, coherence bandwidth and achievable transmission rate. Finally, an overall comparison is made, highlighting similarities and differences taking into account also the conventional (combustion engine) car and the largely investigated in-home scenario.*

## Introduction

The transformation of the conventional vehicles into efficient and intelligent systems requires the deployment of smart electronic components and the establishment of high-speed reliable data links for their interconnectivity. This is typically achieved through dedicated wired data buses, e.g., LIN-bus, CAN-bus. Currently, the car wiring infrastructure is the third factor that contributes to the overall vehicle weight, immediately after the engine and the gearbox. Likewise, in the context of ships, more than 20% of the total weight is due to the electric cables.

The increase of weight has a negative impact on the performance of vehicles and, ultimately, on the energy efficiency. It is therefore of great importance to develop communication technologies that can reduce the amount of wiring. Other than wireless, power line communication (PLC) promises to be a valuable candidate as it can provide robust and reliable communication links by exploiting the existent in-vehicle power delivery infrastructure to convey data. Thus, it can enable the delivery of advanced automotive control applications that need relatively high-speed, flexibility and scalability, or distribute high-speed data streams from internet access, as well as support video streaming from cameras and multimedia and entertainment services. PLC channels are characterized by multipath propagation generated by line discontinuities, due to multiple branches and unmatched loads [1]. This translates into severe frequency selectivity, although, differently from the wireless case, PLC channels are mostly static (no mobility).

The most investigated areas of PLC have been so far related to home networking and communications for the (smart) power grid. Despite the intuitive feeling that PLC can be used for in-vehicle connectivity, no dedicated technology has been developed and not many research results have been published yet. In general, PLC is divided into two classes: narrowband (NB) and broadband (BB). NB PLC operates in the range of frequencies below 500 kHz, and it is used for low rate applications, as command-and-control, remote monitoring and automatic meter reading. BB PLC may signal up to 86 MHz, and it enables streaming high-speed multimedia content. The most recent standards, e.g. HomePlug AV2, are capable of up to 500 Mbps at the physical layer [1]. Such impressive performance is made possible by the adoption of state-of-the-art communication solutions as multi-carrier modulation, multiple-input multiple-output (MIMO) transmission schemes that exploit multiple wires, and powerful turbo codes. A simplified low-power version of the BB HomePlug standard, called HomePlug GreenPHY, can also be used as a valuable alternative to narrowband PLC for low rate applications. In this respect, this standard has been adopted as a reference for communications between the electric vehicle and the charging station.

In-vehicle PLC refers to the set of applications devoted to establish data links inside any means of transportation, i.e., cars, ships, planes, or trains. The investigation of in-car PLC has been documented in [2]-[6]. The in-ship environment is less investigated. Some results are shown in [7]. Moreover, PLC can find application in more unconventional scenarios, such as in planes, in space crafts and in trains, as discussed in [8], [9] and [10], respectively.

In this paper, the focus is on in-car and in-ship PLC. In-car PLC is challenged by the mixture of high channel attenuation, high levels of noise and by low values of the line impedance. The characterization of such quantities is of great importance for the design of optimal transmission techniques. In this respect, an online available database of 193 measurements, performed in a conventional car (CC), i.e. with combustion engine, was analysed in [2]. In-car PLC channel models that follow a top-down and a bottom-up approach were described in [3] and [4], respectively. Noise impairments were modelled in [5] taking into account the periodic components.

Most of the above cited papers deal with the conventional car, equipped with a combustion engine. The electric car has several differences due to a different wiring structure, which is more pervasive due to the need of feeding the electric motors and the presence of a number of electric machines (in particular, current drives and power converters) that inject a great deal of noise. Some studies, based on experimental measurement campaigns, which characterize the PLC channel response and the noise of an electric car (EC), were discussed in [11] and [12]. In [11] the line impedance and the noise introduced by the DC-DC converter of a compact electric car, equipped with a lithium battery pack that provides up to 70 km of autonomy, were additionally analysed. While in [12], measurements of the channel frequency response (CFR) between the vehicle and the external grid during battery charging, together with the symbol error rate evaluation of a CFR simulation model, were also performed.

The aim of this paper is to investigate and assess differences and similarities among the in-car and the in-ship scenarios by exploiting the database of measurements and the initial results reported in [11] and [13], under the same assumptions. The analysis starts from the description of the network structure, of the channel and line impedance properties, as well as of the noise generated by the DC-DC converter and by the motor drives, within the electric car scenario. The results are compared to the findings reported in [2], that refer to a conventional car. Then, the same analysis is performed for the in-ship scenario, reporting the main results obtained by the measurements made in a large cruise ship [13]. The focus is on low voltage BB channels. Moreover, the channel response characteristics are studied in terms of average channel gain (ACG), root-mean-square delay spread (RMS-DS), coherence bandwidth (CB), and maximum achievable rate (i.e. the channel capacity).

Finally, an overall comparison is made considering also the conventional car and the in-home scenario. It will be shown that, despite the differences in network size and topology, as well as in channel characteristics and noise properties, PLC applied to the in-car and the in-ship scenarios can achieve high and similar performance considering the same transmitting power constraints as those typically used in home networks.

**In-Car Scenario**

The in-car scenario is heterogeneous since it comprises vehicles with different powertrains and power grids. It includes electric cars, supplied by batteries, and conventional cars, powered by fossil fuel, whose power grid characteristics differ significantly. Despite the large differences, reliable PLC is possible in both electric and conventional cars, as shown by the results presented in the next sections.

Conventional cars are characterized by a massive wiring infrastructure, with an overall length of several kilometres, but with a limited number of junctions and small single cables length. Furthermore, the return conductor is often missing, since the vehicle chassis is used for it. The noise coming from the powertrain is limited and mainly due to the activity of the spark plugs and the control unit [5], [14].

Electric vehicles are more affected by noise, where the powertrain comprises electric motors controlled by drives. The power grid consists of two sub-networks. The high-power sub-network feeds the motors and it is directly connected to the battery pack. The low-power low voltage sub-network feeds the remaining loads, as the derivation panel, the auxiliary devices and the lights. Both sub-networks are DC, and the low-power sub-network is fed from the high-power sub-network via a DC-DC converter (see Fig. 1). Therefore, the two circuits are not separated and the large noise components generated by the motor drives propagate (and radiate) toward the low-power sub-network, where PLC devices are connected. The resultant noise exhibits both periodic and impulsive components [11].

Fig. 1 shows the power grid scheme and the measurement points of the electric vehicle under test. The vehicle is a best seller in its market, namely the Estrima Birò, a compact 4-wheels electric car manufactured in Italy. The car is equipped with three-phase brushless electric motors, one for each of the two rear wheels. The electric motors are connected to the 48 V high-power sub-network. A DC-DC converter feeds the 12 V low-power sub-network.

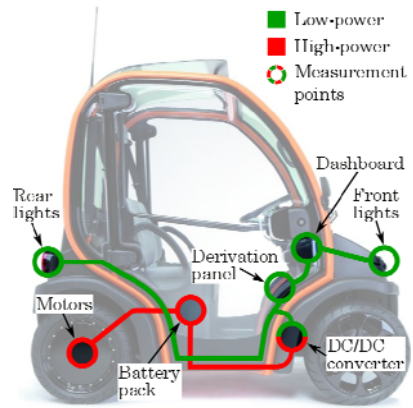

**Figure 1** *Vehicle under test: Estrima Birò. The power distribution grid scheme and the measurement points are also depicted.*

In the 12 V sub-network, the DC-DC converter is connected to the derivation panel, which in turn feeds all the peripherals, as the lights and the dashboard. Therefore, the 12 V sub-network exhibits a star-shape structure, with the derivation panel being the centre. Peripherals are connected to the derivation board through both positive and negative feeding conductors, with nearby devices exploiting the same negative wire. In general, the underlying network deploys cables with length up to 3 meters.

*In-Car Channel Properties*

In this section, the CFR of the 84 measurements obtained from [11] are shown and discussed. The measurements were acquired through a coupler with a flat frequency response up to 100 MHz and using a vector network analyser (VNA) in the 1.8-100 MHz frequency range. During the measurement campaign, two different motor states (on, off) and various equipment activity were considered, as it is detailed in the following. Communications within the low-power sub-network and between the low-power and high-power sub-networks, across the DC-DC converter, are considered. The former case targets the propagation of the PLC signal in the 12 V circuit. In the latter case, there is a lack of electric continuity. In fact, the DC-DC converter can be modelled and acts as a low pass filter, since it is designed for the DC signal conversion only. Thus, the propagation of the broadband high-frequency signal through the DC-DC converter can only be due to coupling and radiated effects.

The analysis of the measurements has shown that the channels can be classified in three different classes, according to the attenuation they exhibit, since this primarily determines the system capacity. However, other classification criteria can be used. The classes make a distinction between low, medium and high-attenuated channels. In general, channels within the 12 V circuit belong to the first two classes, regardless the physical length of the electric paths. Instead, channels across the DC-DC converter belong to the third class, the most attenuated. This result is not surprising since, in this case, the propagation is only due to radiated effects.

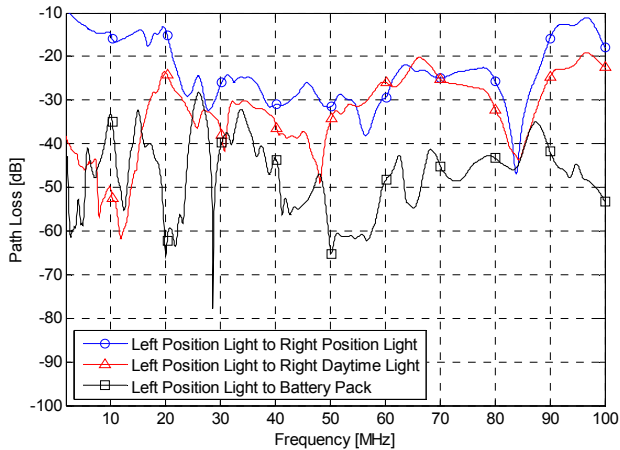

*Figure 2* Path loss of three representative in-car channels [11].

Fig. 2 shows a representative path loss (PL) profile for each class. The PL is defined as $PL(f)=|H(f)|^2$, where $H(f)$ is the CFR. In the figure, the low-attenuated channel refers to the link between the left front position light (FPL) to the right FPL. The medium-attenuated channel refers to the link between the left FPL to the right daytime running light. The high-attenuated channel refers to the link between the left FPL to the battery pack. Interestingly, it can be noted that, despite the large attenuation of the channel across the DC-DC converter, some low-attenuated frequency windows exist. Moreover, although the channels exhibit a moderate selective behavior, due to the reduced multipath propagation into short cables, a low ACG (listed in Table 1) is observed, as later discussed. The PL range and behavior depicted in Fig. 2 are consistent with those reported in [12], although the latter refer to different car states (key ignition positions) and for frequencies up to 30 MHz.

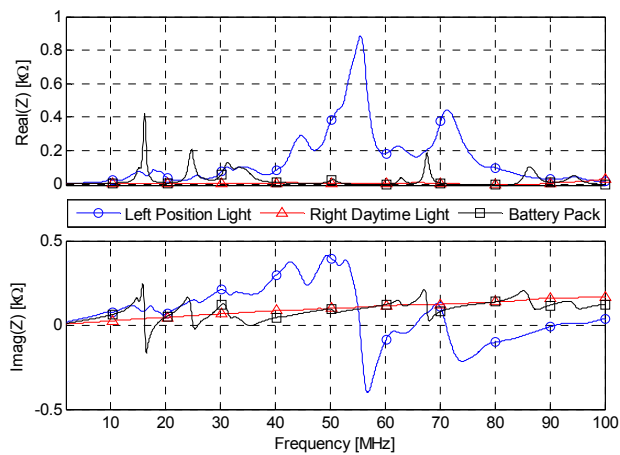

*Figure 3* Impedance of three representative in-car channels [11].

Fig. 3 reports the impedance as a function of frequency. The line impedance is an important quantity as it describes the load seen by the transmitter. Low impedance values make the signal injection challenging. In this respect, the transmission of a broadband high-frequency signal is desirable because the real part of the line impedance reaches significant high levels at middle frequencies. Moreover, the imaginary part of the impedance exhibits an inductive behaviour for frequencies up to 55 MHz and a capacitive behaviour for higher frequencies. The inductive behaviour is explainable considering that light bulbs are built with a metallic wire wounded as a coil. The capacitive behaviour is due to the parasitic capacitance of the small bulbs of the FPL lamps.

*In-Car Noise Properties*

In the following, the noise is discussed focusing on the electric vehicle. Differently from the conventional cars, PLC in electric vehicles is more affected by noise impairments due to the activity of the electric drives.

For the noise experimental acquisition, a digital storage oscilloscope (DSO), with a bandwidth of 1 GHz, was used and connected to each test point highlighted in Fig. 1.

Fig. 4 shows a time-domain noise waveform measured close to the DC-DC converter, on the 12 V sub-network side (see Fig. 1), when the car is in movement with all the equipment switched on.

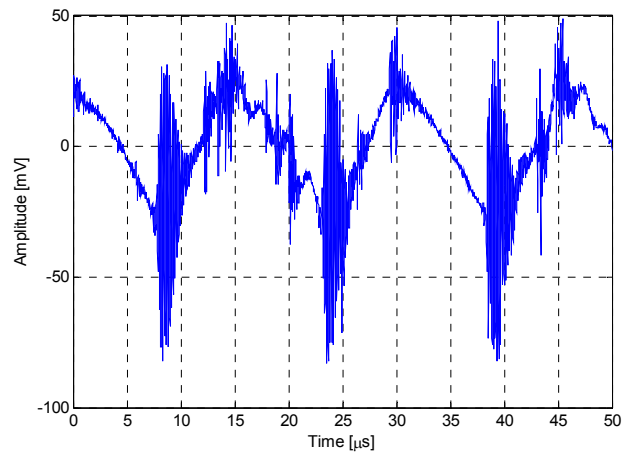

*Figure 4* In-car noise voltage waveform measured at the DC-DC converter (12 V side), with motors and equipment switched on.

As it can be noted, the noise is characterized by a dominant triangular signal plus sharp and fast fluctuations. The former component reflects the switching activity of the DC-DC converter. Basically, it is due to the charge/discharge current which flows in the inductor within the DC-DC converter and is not properly filtered, so that it propagates in the 12 V side of the network.

The sharp fluctuations, instead, are the sum of two contributions, the coupled noise due to the motor drives and the overshooting ripples due to, again, the DC-DC converter. When the car is moving and the motors are running, a large amount of current flows from the battery pack to the motors. Such relevant currents generate a significant magnetic field that concatenates itself with the remaining power plant of the car. This translates into a coupled noise component observable along the ramps. These spikes, induced by the motors, have approximately the same periodic behaviour described in [12], which, however, does not consider the DC-DC converter activity. Instead, the overshooting ripples confined on the switching discontinuities are due to further resonance effects within the DC-DC converter.

The triangular-shape and the coupled noise components (due to the motors) can be analytically described in terms of Fourier series expansions, showing that their effects are limited to the lower frequency range, where the harmonics are concentrated, as detailed in [11]. This is noticeable in Fig. 5 that shows the broadband noise power spectral density (PSD) for two motor states (on, off). The differences are confined at low frequencies. The displayed PSD has been obtained by averaging the noise PSDs over all the measurement points (see Fig. 1) and the considered equipment configurations, as well as for both the motor states.

Contrariwise, the noise contribution due to the overshooting ripples caused by the DC-DC converter translates into a high-frequency component centred around 5 MHz, as shown in Fig. 5. Moreover, the harmonics of these overshooting ripples can be observed in the 15 MHz band. Beside such noise contribution, Fig. 5 highlights a large noise component between 50 MHz and 60 MHz that is not related to the vehicle activity. Rather, it is amenable to coupled RF signals that were present in the laboratory site during measurements. Likewise, the PSD components in the range 30-40 MHz and above 87 MHz are due to amateur and broadcast radio transmissions, respectively.

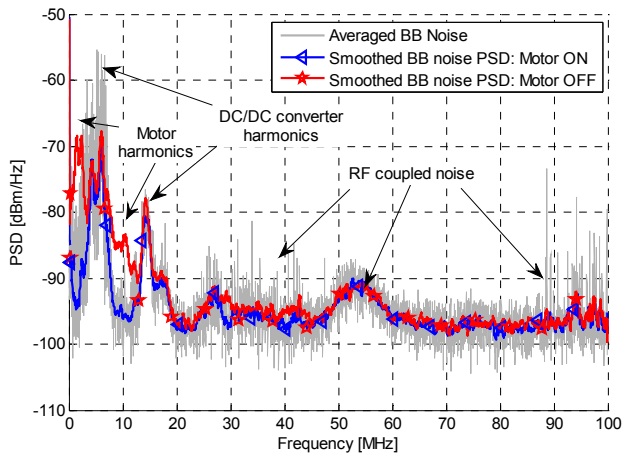

*Figure 5* Broadband in-car noise PSD for two motor states. The average PSD value is also shown.

Conventional cars show a noise PSD tens of dB lower than the one experimented in the electric vehicle described in this section, as reported for instance in [14].

**In-Ship Scenario**

In this section, the attention is turned to the in-ship environment, where the main results obtained from the measurement campaign carried out in a cruise ship [13] are discussed.

*In-Ship Channel Properties*

A total of 92 time-domain measurements were carried out using a signal pulser together with a DSO (with a sampling rate of 200 Msamples/s), connected via a capacitive coupler having a CFR with a flat behaviour up to 50 MHz. The vehicle under test was the 116,000 tonnes cruise ship depicted in Fig. 6. In this scenario, the power is supplied by asynchronous diesel generators via medium voltage (MV) lines to MV/LV substations and distributed to the decks through low voltage (LV) three-phase wires with non-distributed neutral. In detail, the power is vertically distributed from the substation switchboard (SS) to all decks, reaching the distribution boards (DBs) and deploying cables ranging from 80 to 100 meters. Each DB is then connected by a bus-bar, of length around 40 meters, to the room service panels (RPs) that serve a small number of rooms each [13]. Therefore, the network topology is the ensemble of two differently structured sub-networks: a star-style network from the SS to the DBs and a bus-style network from the DB to the RPs. The channels referred to these sub-networks are named as SS-DB and DB-RP, respectively. Since the underlying structure is completely different, these two sub-networks are separately considered in the rest of this section.

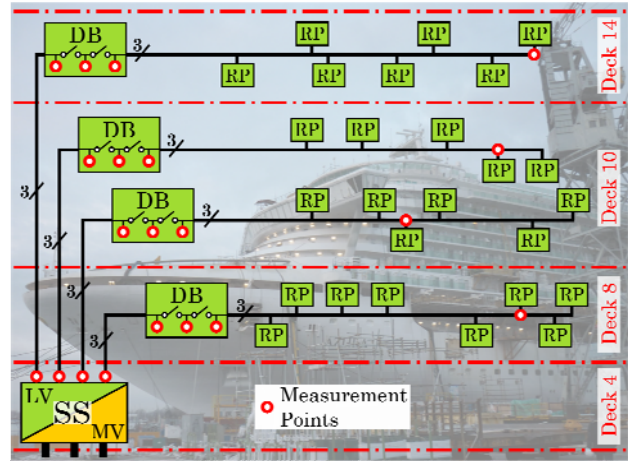

*Figure 6* Power distribution grid scheme of the cruise ship Azura, built by Fincantieri. The measurement points are also shown.

The measurements in [13] concern only the LV distribution network, as depicted in Fig. 6. Since there are three available conductors (denoted with R, S, and T), a 2×2 MIMO channel can be established transmitting and receiving among the R and T conductors, as well as between the S and T conductors, in a differential mode. For the measurements, the two direct links and the two cross-links responses were measured using a single-input single-output (SISO) configuration and closing the other two ports on a known 50 Ω load.

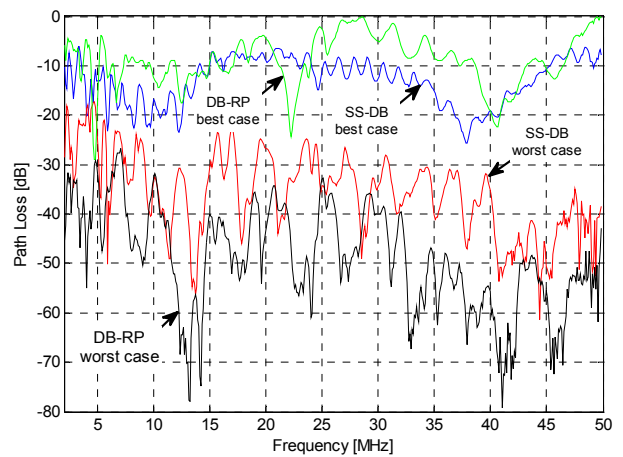

*Figure 7* Best and worst path loss profile for the two in-ship power sub-networks (i.e., SS-DB and DB-RP).

The best and the worst path loss profiles of the channels belonging to the two sub-networks are reported in Fig. 7. Due to the coupler CFR and to the measurement procedure, the frequency range is limited to 1.8-50 MHz. As it can be noted, the PL ranges between -10 and -60 dB (as for the in-car scenario previously discussed), despite a larger network with much longer cables. However, the ACG listed in Table 1 is higher than in the electric car scenario. This is due to the limited amount of branches in the SS-DB sub-networks and to a better quality of the deployed cables. Frequency selectivity is higher than that observed in the in-car scenario (as stated by the low average CB level listed in Table 1), although for the best PL profile it is quite small. This is because such channels are associated to cable links without derivations, where the reflection effects are less pronounced.

*In-Ship Noise Properties*

Noise measurements were carried out in a number of points in order to evaluate the PSD using a spectrum analyser in the band 1.8-50 MHz. Two states of the ship were considered; with switched on or off diesel generators. Small variations of PSD levels along frequency were found, with a floor ranging between -135 dBm/Hz and -145 dBm/Hz, depending on the node position. It has to be said that the ship was not in navigation conditions. In such a case, the experienced noise may be higher due to load and people activity. Thus, a noise PSD with a conservative floor value of -110 dBm/Hz can be considered, as it will be done in the performance analysis section. This value appears to be an appropriate compromise between the -95 dBm/Hz measured in [7] and the nearly -132 dBm/Hz floor observed in [15].

## Statistical Analysis & Performance

In order to statistically assess and characterize the PLC channels in the considered scenarios, in this section, the most commonly used metrics, i.e. ACG, RMS-DS, CB and the achievable capacity, are defined and analysed.

*Metric Definitions*

The ACG provides a measure of the average channel attenuation with frequency and it is defined as

$$ACG = \frac{1}{f_2 - f_1} \int_{f_1}^{f_2} |H(f)|^2 df,$$

with $H(f)$ being the CFR in the frequency range $[f_1, f_2]$. The energy dispersion of the channel impulse response $h(t)$, with duration $D$, is measured with the RMS-DS, named $\sigma_\tau$, computed as

$$\sigma_\tau = \sqrt{\int_0^D \tau^2 \frac{|h(t)|^2}{\int_0^D |h(w)|^2 dw} d\tau - \left(\int_0^D \tau \frac{|h(t)|^2}{\int_0^D |h(w)|^2 dw} d\tau\right)^2}.$$

Whereas, the coherence bandwidth at a specified level $\rho$, namely $B^\rho$, yields the frequency at which the correlation function $R(v)$ falls below $\rho$ times its maximum value. Thus, $B^\rho$ is such that $|R(B^\rho)|=\rho|R(0)|$, with

$$R(v) = \int_{f_1}^{f_2} H(f+v)H^*(f)df.$$

In this paper $\rho=0.9$ is considered. Finally, the achievable capacity $C$, under the assumption of colored Gaussian background noise, is computed as

$$C = \int_{f_1}^{f_2} \log_2\left(1 + \frac{P_{tx}(f)|H(f)|^2}{P_w(f)}\right) df,$$

where $P_{tx}(f)$ and $P_w(f)$ are the transmitted signal and the noise PSD at frequency $f$, respectively. The achievable capacity forecasts the maximum transmission rate that can be sustained by a communication channel. The formula can be easily extended to the MIMO case [13].

*In-Car and In-Ship Comparison*

In this section, the aim is to compare the in-car and the in-ship scenarios, assessing the dependencies and evaluating the average values of the previously discussed performance metrics. In order to perform a fair comparison, in the following the considered frequency range is 1.8-50 MHz, unless otherwise stated.

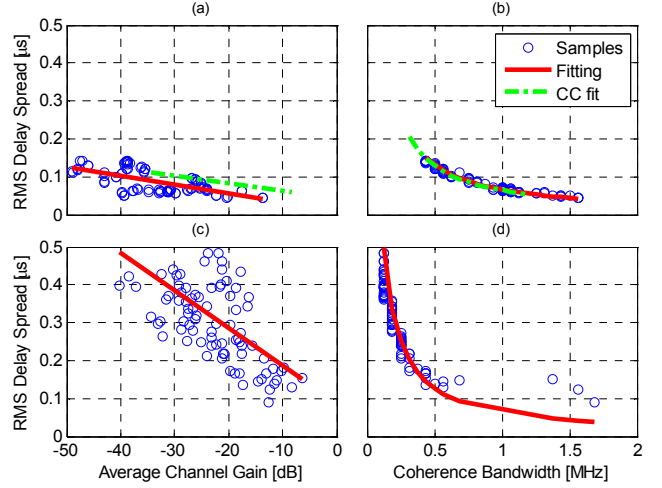

*Figure 8* Comparison among the electric car, (a) and (b), and the in-ship, (c) and (d), scenarios. The CC measurements fit [2] is also depicted.

In Fig. 8, the RMS-DS is analyzed w.r.t. the ACG (on the left side) and the CB (on the right side) for two different scenarios, in-car (on top) and in-ship (on bottom). The fit of the CC measurements discussed in [2] is also depicted. The scatter plots (blue circles) correspond to the measurements. As it can be noted, the RMS-DS and the ACG are negatively related, as shown by the robust regression fit (red line). In particular, a slightly lower attenuation is exhibited by the conventional car compared to the electric car, as later clarified in Table 1. The slope of the robust fit line for the in-ship scenario (Fig. 8c) is roughly five times the slope of the in-car scenario (Fig. 8a). In addition, the in-car scenario experiences lower delay spread, but higher attenuation, since values are more shifted towards the left bottom corner, w.r.t. the in-ship case. This is even more noticeable looking at the average metrics reported in Table 1, later described. Furthermore, on the right side of Fig. 8, it can be noted as the coherence bandwidth is inversely related to the RMS-DS, as shown by the hyperbolic fit (red curve).

Now, the focus is turned into the performance evaluated in terms of achievable capacity. Fig. 9 shows the capacity complementary cumulative distribution function (CCDF) computed using the electric car channel and noise measurements (labeled with EC). As a further term of comparison, the capacity CCDF of the conventional car (labeled with CC) in the 1.8-50 MHz, together with the capacity CCDF in the extended 1.8-100 MHz band for the EC, are also shown. Fitting curves are also displayed. The capacity is computed assuming a transmitted signal PSD of -50 dBm/Hz and the measured noise PSD in Fig. 5 for the electric car. For the conventional car, a flat noise PSD level of -120 dBm/Hz is considered [2]. It should be noted that the conventional car outperforms the electric car since the motor states (on, off) affect the latter and higher noise has been observed. Nevertheless, high link capacity is achievable. This is even more prominent when considering the 1.8-100 MHz frequency range for the electric car

(no measurements beyond 50 MHz are available for the conventional car). In fact, for the electric car, with probability 0.9 the capacity can exceed 500 Mbps in the 1.8-100 MHz band, which doubles the value of 220 Mpbs in the 1.8-50 MHz band.

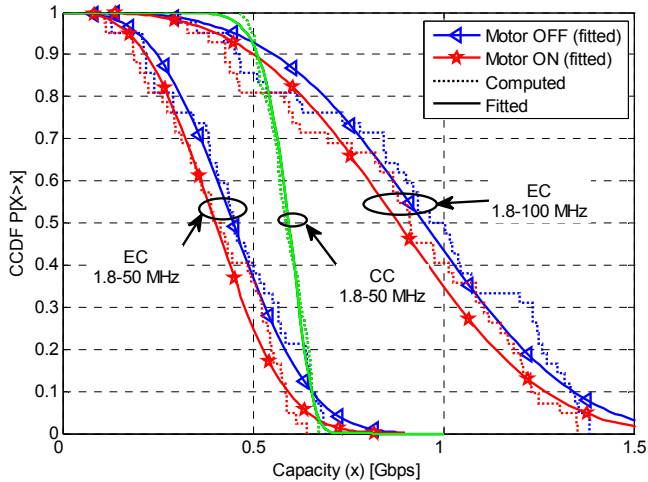

*Figure 9* Electric car PLC performance in terms of capacity CCDF for 1.8-50 MHz and 1.8-100 MHz frequency ranges. The capacity CCDF for the CC in the 1.8-50 MHz band is also shown.

Conversely, the capacity CCDF obtained considering the in-ship channel measurements is reported in Fig. 10. It has been computed assuming a transmitted signal with a PSD of -50 dBm/Hz and an additive white Gaussian noise with a PSD equal to -110 dBm/Hz (which is a conservative floor value, as previously discussed). Since three phases are available, the capacity is estimated for the PLC links that use the same phase (direct links), and the links that use different phases at the transmitter and receiver (coupled links). It can be noted, in Fig. 10, as the capacity of the direct links outperforms that of the coupled ones, for both the considered sub-networks. Moreover, if MIMO transmission is used (under the same SISO assumptions and dividing the available power equally among all the possible transmitting modes, namely two, as discussed in [13]), the capacity will have the potentiality to almost double, compared to the SISO case.

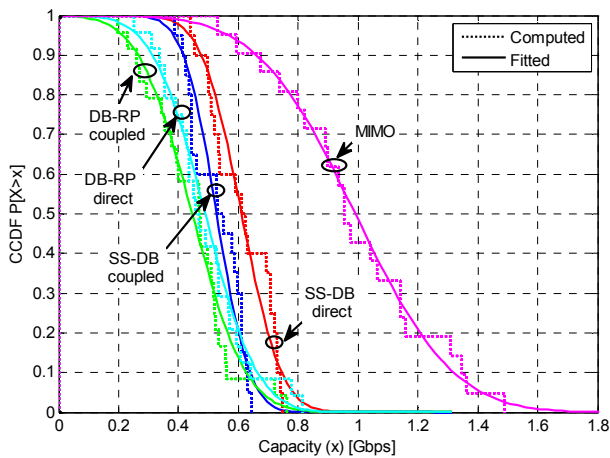

*Figure 10* Capacity CCDF for the two different in-ship sub-networks (in SISO configuration) and for the 2×2 MIMO transmission, both in the 1.8-50 MHz band.

Table 1 summarizes the average value of the performance metrics computed from the measurements in each considered scenario (where $\overline{X}$ indicates the average) in the bandwidth 1.8-50 MHz. As a further comparison, the metric values for the more popular in-home case (considering the 1300 measurements discussed in [16]) and for the online available database of 193 measurements analysed in [2] (which concerns a conventional car), are also listed.

| Scenario | Bandwidth (MHz) | $\overline{ACG}$ (dB) | $\overline{DS}$ (μs) | $\overline{CB}$ (kHz) | $\overline{C}$ (Mbps) |
|---|---|---|---|---|---|
| In-home [16] | 1.8–50 | −33.10 | 0.357 | 251.48 | 539.11 |
| In-car (EC) | 1.8-50 | −33.25 | 0.086 | 874.26 | 425.67 |
| In-car (CC) | 1.8-50 | −27.33 | 0.102 | 677.14 | 589.97 |
| In-ship | 1.8–50 | −22.89 | 0.320 | 258.83 | 511.23 |

*Table 1* Average statistical channel metrics for different scenarios.

It is interesting to note the following. Firstly, the in-ship channels exhibit the higher ACG. Secondly, both in-car scenarios have an ACG similar to the in-home case. Thirdly, the high RMS-DS ($\sigma_\tau$) value and the low CB ($B^{0.9}$) value for the in-ship case, caused by the large amount of branches in the DB-RP sub-networks, are very similar to those of the in-home scenario. Moreover, the lower attenuation exhibited by the in-ship channels is due to the limited amount of branches in the SS-DB sub-networks, to a better cables quality and to a simpler topology, despite the length. Conversely, the lower delay spread and higher coherence bandwidth of the in-car channels is due to the low multipath propagation into short cables length. While, the high attenuation in the EC is due to poor electric wiring characteristics, to the low line impedance values (see Fig. 3) and to the attenuation introduced by the DC-DC converter.

Finally, looking at the average capacity reported in Table 1, it can be noted as the considered scenarios exhibit similar performance (of about 500 Mbps), despite the different network structure, the channel properties and the noise sources. The average capacity for the electric car scenario is slightly lower than the others. This is mainly due to the strong effect of the measured background noise (see Fig. 5).

## Conclusions

The power line communication channel in the in-vehicle scenario has been discussed. In particular, two environments have been considered, the in-car case (particularly, the electric car) and the in-ship case. It has been shown that, in both cases, PLC has the potentiality to offer high-speed data connectivity, despite the different nature of the underlying network structure and background noise properties. The measurements analysis has shown that, although the in-car network has a compact size, the PLC channel is affected by frequency selective fading and high attenuation. The in-ship channels exhibit (in average) less attenuation since there are fewer discontinuities and better cables, although they are significantly longer. Moreover, the background noise of the electric car has been analysed, showing a prominent and severe influence at low frequencies, especially where NB PLC operates. This is primarily due to the switching activity of the DC-DC converter, together with the electric drives for the motors. The DC-DC converter noise affects also the BB PLC spectrum, in particular at 5 MHz and at its harmonics. The measurements and the literature have shown for the in-ship scenario a noise PSD (in the 1.8-50 MHz) with a floor ranging from -95 to -145 dBm/Hz. Moreover, the presence of a three-phase distribution network allows the usage of MIMO transmission techniques that have the potentiality of almost doubling the channel capacity. Similar capacity improvements can be obtained in the in-car scenario by extending the transmission band to 1.8-100 MHz.

**Acknowledgements**

The Authors wish to thank Estrima, partner of the project ESTAMOS, that made available the electric car for the field trials and Fincantieri, partner of the project "On board systems with PLC technology" for giving access to the Azura ship.

**Alberto Pittolo** received from the University of Udine, Italy, the Laurea degree in electrical engineering (2009) and the Laurea Specialistica degree (2012) in electrical and telecommunications engineering, with honours. He is currently pursuing the Ph.D at the University of Udine. His research interests and activities are channel modelling, physical layer security and resource allocation algorithms for both wireless and power line communications.

**Marco De Piante** received the Laurea degree in electrical engineering (2007), the Laurea Specialistica degree in electrical and telecommunications engineering (2010) and the Ph.D (2014) from the University of Udine, Italy. He is currently a postdoctoral researcher at the University of Udine. His research interests and activities include infrared communications, RF electronics and PLC communications.

**Fabio Versolatto** (GSM'10) received the Laurea degree (2007) and the Laurea Specialistica degree (2009) in electrical engineering (both summa cum laude), and the Ph.D degree in industrial and information engineering (2013), from the University of Udine, Italy. His research interests are in the field of channel modelling for power line communications and digital communication algorithms. He received the best student paper award at the IEEE Int. Symp. on Power Line Commun. and Its App. 2010.

**Andrea M. Tonello** (M00, SM12) was with Bell Labs, Lucent Technologies in the Advanced Wireless Technology Laboratory, Whippany, NJ from 1997 to 2002 as MTS, Technical Manager and Managing Director. From 2003 to 2014 he has been a researcher and an associate professor at the University of Udine, Italy where he founded the WiPli Lab. From 2015, he is the chair of the Embedded Communication Systems group at the University of Klagenfurt, Austria. He is also the founder of the spin-off company WiTiKee. He holds the laurea degree (1996) and the PhD (2002) in electrical engineering. He is a Distinguished Lecturer of the IEEE Vehicular Technology Society since 2011 and he is the chair of the ComSoc TC on PLC. He received several awards and he covers-ed several editorial roles in IEEE journals (TVT, TCOM, ACCESS).